\documentclass[preprint,12pt]{elsarticle}
\usepackage[english]{babel}
\usepackage[utf8]{inputenc}
\usepackage[T1]{fontenc}
\usepackage{newtx}
\usepackage{graphicx}
\usepackage{amsmath}
\usepackage{xcolor}
\usepackage[colorlinks,allcolors=blue]{hyperref}
\usepackage{xspace}
\usepackage{enumitem}
\usepackage{caption}
\usepackage{lineno}

\setlist{noitemsep, nosep}
\setlist[enumerate,1]{label = (\roman*)}

\newcommand{\MeV}{\mbox{Me\kern-0.15em V}\xspace}
\newcommand{\GeV}{\mbox{Ge\kern-0.15em V}\xspace}
\newcommand{\A}{\textit{A}\xspace}
\newcommand{\GeVc}{\mbox{\GeV{}\kern-0.15em/\kern-0.05em\textit{c}}\xspace}
\newcommand{\MeVc}{\mbox{\MeV{}\kern-0.15em/\kern-0.05em\textit{c}}\xspace}

\newcommand{\AGeVc}{\mbox{\A{}\,\GeVc}\xspace}

\newcommand{\pt}{\ensuremath{p_\mathrm{T}}\xspace}
\newcommand{\Dpt}{\ensuremath{|\Delta\boldsymbol{\pt}|}\xspace}

\begin{document}

\begin{frontmatter}

\title{Simple Power-law Model for generating correlated particles\tnoteref{grant}}

\tnotetext[grant]{Supported by Polish National Science Centre grant 2018\slash 30\slash A\slash ST2\slash 00226}

\author[ujk,ncbj]{Tobiasz Czopowicz}
\ead{tobiasz.czopowicz@ncbj.gov.pl}
\address[ujk]{Institute of Physics, Jan Kochanowski University, Kielce, Poland}
\address[ncbj]{National Centre for Nuclear Research, Otwock/Warsaw, Poland}

\begin{abstract}
A search for the critical point of the strongly interacting matter by studying power-law fluctuations within the framework of intermittency is ongoing. In particular, experimental data on proton and pion production in heavy-ion collisions are analyzed in transverse momentum space.
In this regard, a simple Monte Carlo model with an explicit power-law multi-particle correlation in transverse momentum space is introduced. The model is intended as a phenomenological tool to study the sensitivity of intermittency analyses to power-law correlated particles in the presence of various detector effects.
\end{abstract}


\end{frontmatter}

\newpage
\noindent
{\bf PROGRAM SUMMARY}

\begin{small}
\noindent
{\em Program Title:} Power-law Model \\
{\em CPC Library link to program files:} (to be added by Technical Editor) \\
{\em Developer's repository link:} (if available) \\
{\em Code Ocean capsule:} (to be added by Technical Editor)\\
{\em Licensing provisions(please choose one):} CC BY 4.0 \\
{\em Programming language:} ANSI C \\
{\em Supplementary material:} \\
{\em Nature of problem (approx. 50-250 words):}\\
Generating particles (products of a collision) with correlations in tranverse momentum
expected for the QCD critical point \\
{\em Solution method (approx. 50-250 words):}\\
Generating momenta of particles with a power-law correlation in transverse momentum while
preserving a given transverse-momentum inclusive distribution \\
{\em Additional comments including restrictions and unusual features (approx. 50-250 words):}\\
\end{small}

\newpage

\section{Motivation}
\label{sec:motivation}

One of the goals of high-energy heavy-ion physics is to locate the critical point (CP) in the
phase diagram of the strongly interacting matter. Theoretical studies suggest a smooth crossover
transition at small baryochemical potential $\mu_\mathrm{B}$ and high temperature $T$~\cite{Aoki:2006we}.
At lower $T$ and larger $\mu_\mathrm{B}$, a first-order phase transition is expected~\cite{Asakawa:1989bq}.
The CP is a hypothetical end point of the first-order phase transition with properties of the
second-order phase transition.

In the vicinity of CP, fluctuations of the order parameter become self-similar~\cite{Antoniou:2006zb}, 
belonging to the 3D-Ising universality class. This can be detected by studying particles' fluctuations
in the transverse momentum, \pt, space within the framework of intermittency analysis by use of Scaled
Factorial Moments (SFM). A search for such power-law correlations was proposed in
Refs.~\cite{Bialas:1985jb, Bialas:1988wc, Satz:1989vj, Gupta:1990bi} and experimental data on proton
and pion multiplicity fluctuations have been analyzed in transverse momentum space~\cite{Anticic:2009pe,
Anticic:2012xb, Davis:2020fcy}.

To study the sensitivity to detect power-law correlated particles in the presence of various
detector effects, a simple Monte Carlo model that can generate particles with phenomenological properties expected near the CP was developed.
The software implementation is introduced in Sec.~\ref{sec:model}, followed by example results in Sec.~\ref{sec:results}.

\section{Power-Law Model}
\label{sec:model}

The Power-Law Model generates a given number of events with particles. Number of particles for each event
is drawn from a given multiplicity distribution and each particle's transverse momentum from a given
transverse momentum distribution.
Additionally, for a given fraction of particles, power-law correlation in transverse momentum is
introduced. Results (events with list of particles' momenta components) are stored in a text file
and can be used for calculating SFM or undergo further processing (e.g. momentum smearing to
mimic detector's momentum resolution).
The model is written in ANSI C with no external dependencies.
It uses SFC64~\cite{sfc64} algorithm for (pseudo)randomness.

\subsection{Power-law correlation}
\label{sec:correlation}

The model allows for generating groups of correlated particles (pairs, triplets, quadruplets, etc.).
These correlations are introduced using the average, over particles in a group, pair transverse momentum
difference $S$. For a given number of particles in a group $g$ that form
$g_{\mathrm{p}} = \binom{g}{2} = g(g-1)/2$ pairs, $S$ is defined as
\begin{equation}
  S = \frac{1}{g_{\mathrm{p}}}\sum\limits_{i=1}^{g}\sum\limits_{\substack{j=i+1}}^{g}\Dpt_{i, j}\,,
\end{equation}
where $\Dpt_{i, j} = \sqrt{(p_{x,i} - p_{x,j})^{2} + (p_{y,i} - p_{y,j})^{2}}$ denotes vector difference of transverse momenta of $i$-th and $j$-th particles and summation runs over all $g_{\mathrm{p}}$ pairs of $g$ particles.
For a correlated pair, when $g=2$ and $g_{\mathrm{p}}=1$, $S$ simply equals the difference of the
transverse momenta of the two particles, $S=\Dpt$.

The correlation is introduced by generating $S$ according to the power-law distribution with a given exponent $\phi$:
\begin{equation}\label{eq:rhoS}
  \rho_{S}(S) = S^{-\phi}\,.
\end{equation}
Due to the imposed scale-invariance of the power-law distribution, $S$ for sub-groups of $g$ particles follow the same power law (e.g. quadruplet of particles correlated with $\phi$ contains 4 triplets correlated with $\phi$ and 6 pairs also correlated with $\phi$).

\subsection{Event multiplicity}
\label{sec:multiplicity}

Number of events, $N_{\text{events}}$, is one of the input (command-line) parameters. Processing will stop
after reaching the requested value.
The number of particles in each event is drawn from either a standard distribution (Poisson with a given
expected value or Gaussian with a given mean and standard deviation) or a custom distribution supplied
in a text file. Event multiplicity can also be set to a constant value.

\subsection{Transverse momenta components}
\label{sec:pxpy}

To generate transverse momentum components of each particle, the following parameters are used:
\begin{enumerate}
  \item desired ratio of total number of correlated particles to all particles $r_{1}$ \mbox{(default: 0.5)},
  \item power-law exponent $\phi$ \mbox{(default: 0.65)},
  \item minimum and maximum value of the average pair momentum difference $S$ of correlated particles
    \mbox{(default: $S_{\text{min}} = 1^{-15}$ \GeVc, $S_{\text{max}} = 1.2$ \GeVc)},
  \item number of correlated particle in group $g$ \mbox{(default: 2)},
  \item single-particle transverse momentum distribution $\rho_{\pt}(\pt)$ in a text file
    (default:
    \begin{equation}
    \rho_{\pt}(\pt) = \pt e^{-6\pt}
    \end{equation}
    for \mbox{$0 \leq \pt \le 1.5$~\GeVc)}.
\end{enumerate}

The decision wether to generate an uncorrelated particle or a correlated group of particles is based on the fraction of correlated particles already generated. If generating a new correlated group would cause its value to exceed the value of the input parameter $r_{1}$, an uncorrelated particle is produced; otherwise, correlated group is chosen.

\subsubsection{Uncorrelated particles}

Generating uncorrelated particle's transverse momentum components takes the following steps:
\begin{enumerate}
  \item draw \pt form the supplied transverse momentum distribution $\rho_{\pt}$,
  \item draw azimuthal angle $\phi$ from a uniform distribution [$0,2\pi$),
  \item calculate the components, as $p_{\mathrm{x}} = \pt\text{cos}(\phi)$ and $p_{\mathrm{y}} = \pt\text{sin}(\phi)$.
\end{enumerate}

\subsubsection{Correlated particles}

Generating transverse momentum components of correlated particles requires \emph{a priori} preparation of:
\begin{enumerate}
  \item a histogram of transverse momentum values $H_{\pt}$ for all correlated particles, and
  \item an array of average pair momentum differences $S_i$ for all correlated groups.
\end{enumerate}
The total number of correlated particles to be generated is estimated as
\begin{equation}
  n_{\mathrm{c}} = N_{\text{events}} \cdot \langle N\rangle \cdot r_{1}\,,
\end{equation}
where $\langle N\rangle$ is the mean value of the requested multiplicity distribution. Than, the total number of correlated groups to be generated is
$$
  n_{g} = n_{\mathrm{c}}/g\,.
$$

For each of the $n_{\mathrm{c}}$ correlated particles to be generated, their transverse momenta values are randomly drawn from the input distribution $\rho_{\pt}$ and fill the $H_{\pt}$ histogram. This histogram will be used for obtaining transverse momenta of correlated particles.

An array of $n_{g}$ values of $S$ following the power-law distribution $\rho_{S}$ from
Eq.~\ref{eq:rhoS} is computed. Each value is calculated using the inverse transform sampling method
(with additional constraints on minimum and maximum values of $S$) as
\begin{equation}
  S_{i} =
    \begin{cases}
      \big[(S_{\text{min}})^{1-\phi} + \left((S_{\text{max}})^{1-\phi} -
      (S_{\text{min}})^{1-\phi}\right)R_{i}\big]^{\frac{1}{1-\phi}} & \text{for}~\phi \neq 1\\[2ex]
      S_{\text{min}}\cdot\left(\frac{S_{\text{max}}}{S_{\text{min}}}\right)^{R_{i}} & \text{for}~\phi=1
    \end{cases}\,,
\end{equation}
where $i = 1\ldots n_{g}$ and $R_{i}$ are (pseudo)random numbers from a uniform distribution in [0,1).
Particles in groups with large $S$ tend to have higher \pt and are easier to generate at earlier stage. Therefore, the $S_{i}$ array is sorted in descending order.

Whenever a correlated group of $g$ particles is to be generated, the next value from the $S_{i}$ array is used. A temporary probability distribution (histogram) of possible values of the average \pt from $H_{\pt}$ is prepared, and one value is drawn.

In the $p_\mathrm{x}-p_\mathrm{y}$ plane, correlated particles from the group are evenly positioned on a circle with diameter $S_{i}$, centered at their average \pt. Then, components of their transverse momenta are calculated and stored.

As the last step, histogram $H_{\pt}$ is updated by removing the used values of \pt of all $g$ particles.

This geometric construction provides a simple and controlled way to impose a fixed average pair momentum difference without modeling physical emission mechanism.

\subsection{Particles' longitudinal momentum components}
\label{sec:pz}

Both correlated and uncorrelated particles' longitudinal momentum components $p_\mathrm{z}$ are calculated
independently from \pt from a uniform center-of-mass rapidity distribution. The following parameters are used:
\begin{enumerate}
  \item minimum and maximum value of rapidity in the center-of-mass frame
    (default: \mbox{$y_{\text{min}}^{\text{CMS}} = -0.75$,} {$y_{\text{max}}^{\text{CMS}} = 0.75$}),
  \item mass of particles $m$ in $\GeVc^2$ (default: 0.938, proton mass),
  \item rapidity of the center-of-mass in laboratory frame
    \mbox{(default: $y_{\text{cms}}^{\text{LAB}} = 2.88$}, value for fixed-target collision with beam momentum $p_\text{beam} = 150\AGeVc$)
\end{enumerate}

The center-of-mass rapidity distribution is assumed to be flat in a given range
($y_{\text{min}}^{\text{CMS}};y_{\text{max}}^{\text{CMS}}$) and one value $y^{\text{CMS}}$ is chosen at
random. Using a given particle mass and generated \pt, transverse mass is calculated as
\begin{equation}
  m_\mathrm{T} = \sqrt{(\pt)^{2}+ m^{2}}\,.
\end{equation}

Knowing the rapidity of the center-of-mass in the laboratory frame $y_{\text{cms}}^{\text{LAB}}$ and transverse mass allows us to calculate
$p_{z}$:
\begin{equation}
  p_\mathrm{z} = m_\mathrm{T} \cdot \text{sinh}(y^{\text{CMS}} + y_{\text{cms}}^{\text{LAB}})\,.
\end{equation}

\section{Model performance and results}
\label{sec:results}

\begin{figure}[!htb]
  \includegraphics[width=.33\textwidth]{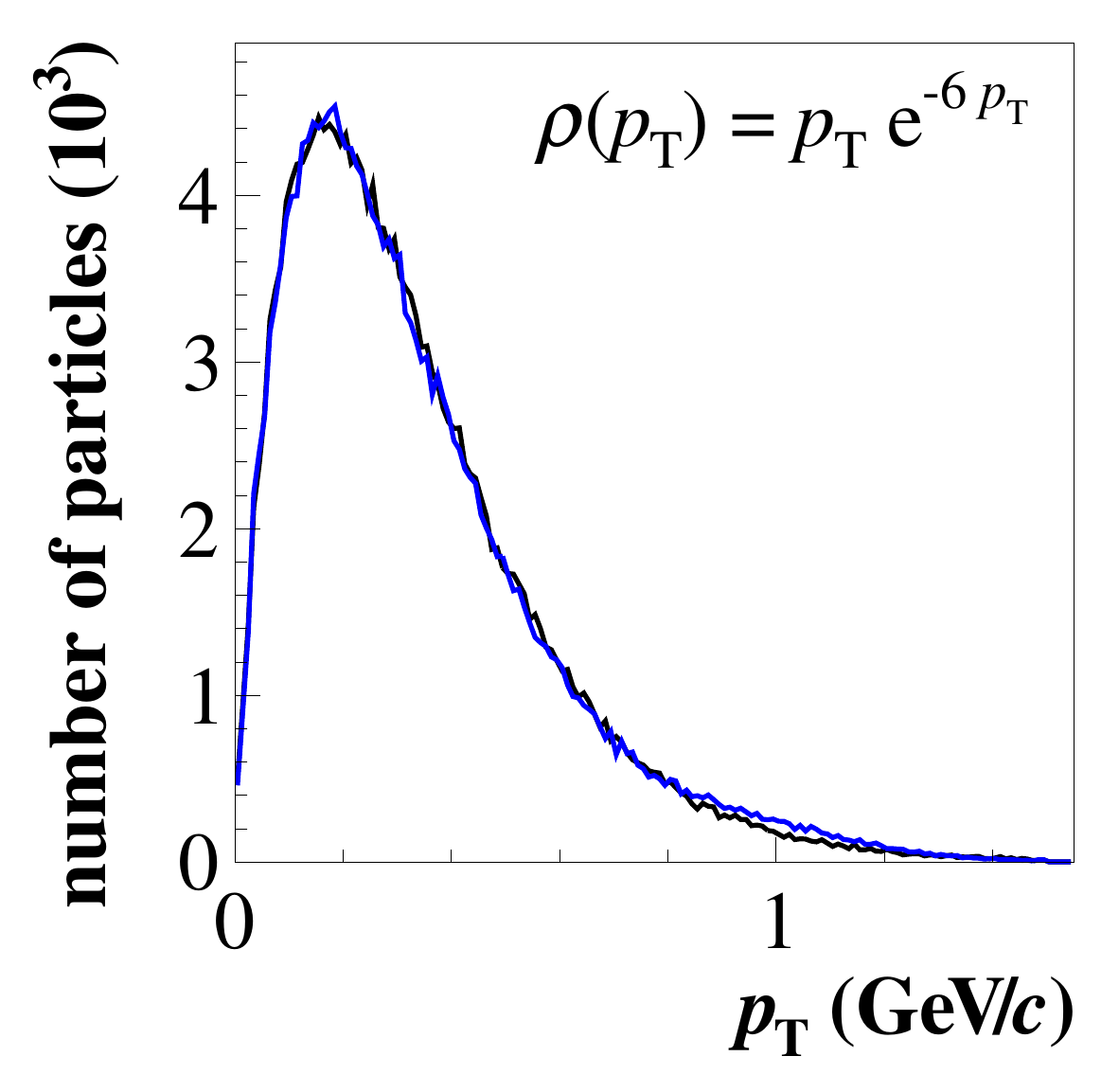}\hfill
  \includegraphics[width=.33\textwidth]{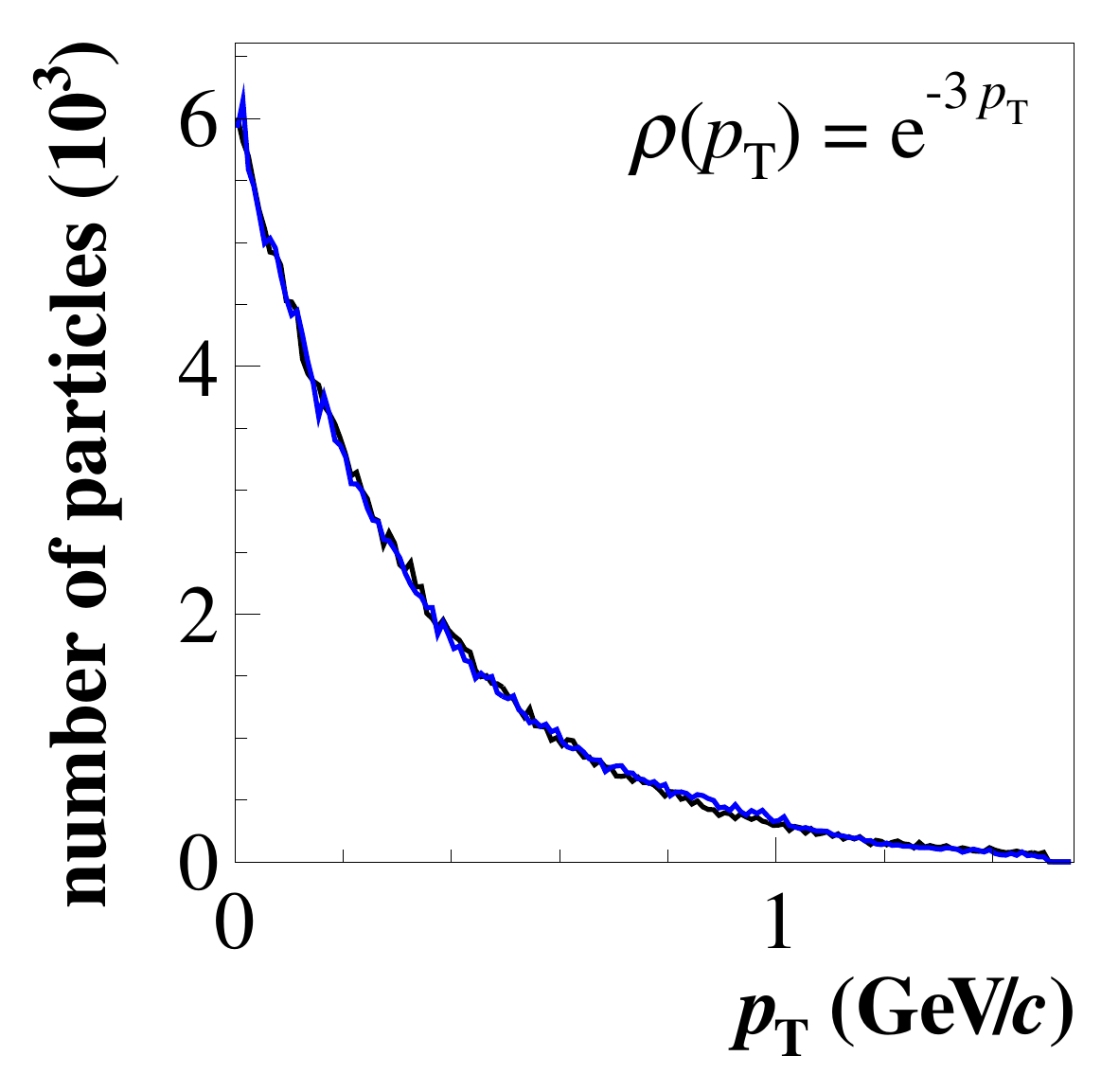}\hfill
  \includegraphics[width=.33\textwidth]{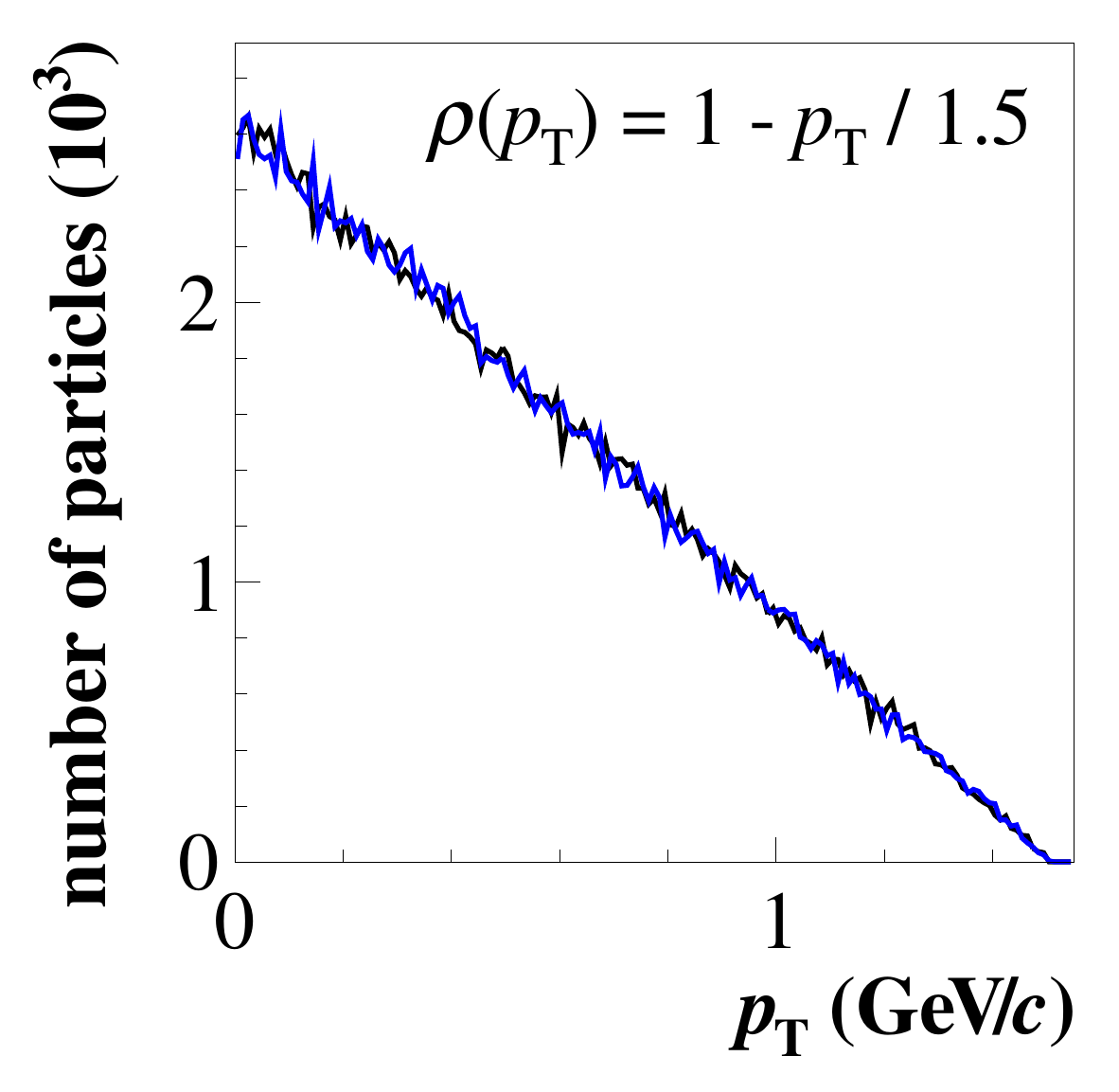}\\
  \includegraphics[width=.33\textwidth]{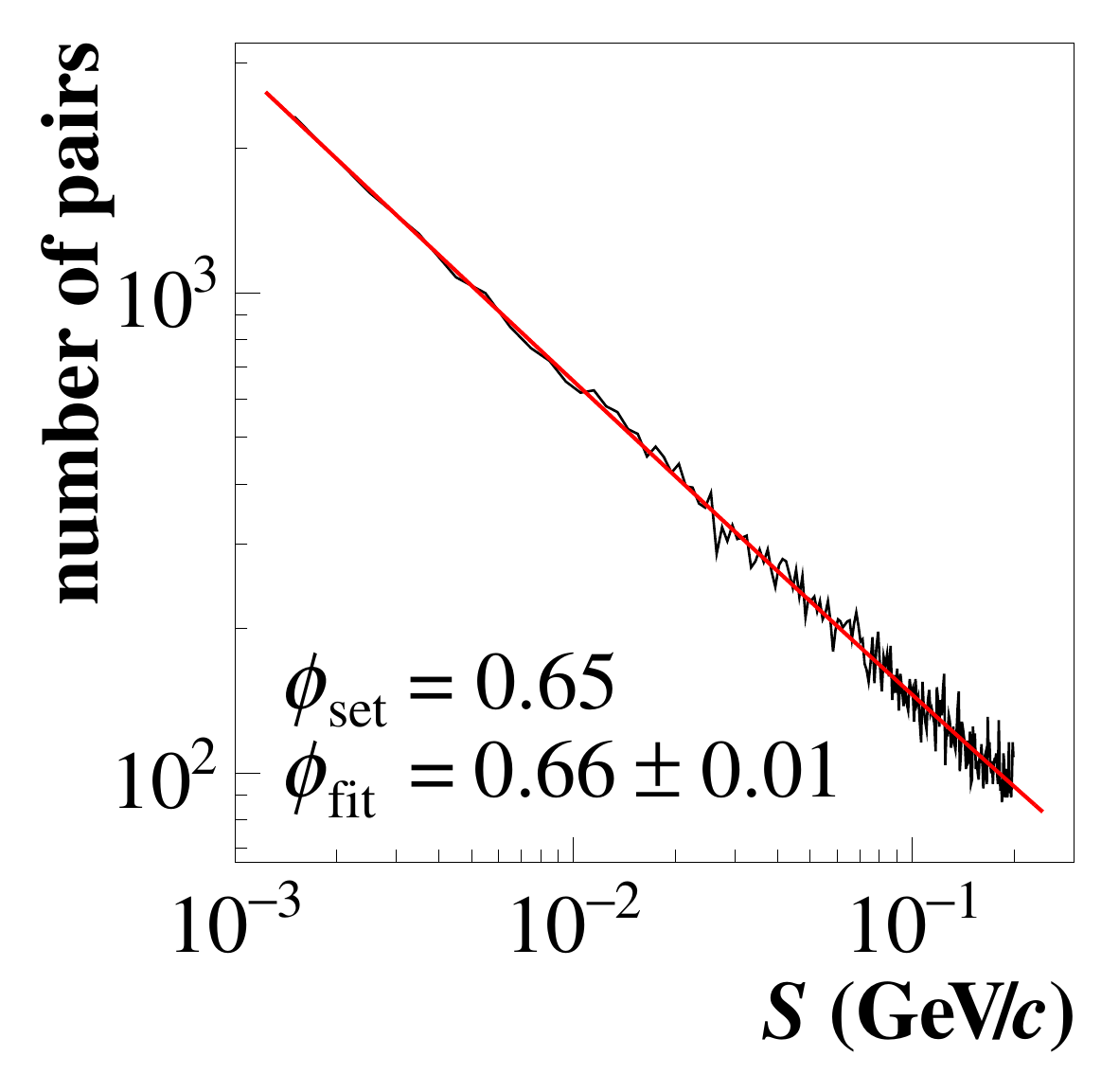}\hfill
  \includegraphics[width=.33\textwidth]{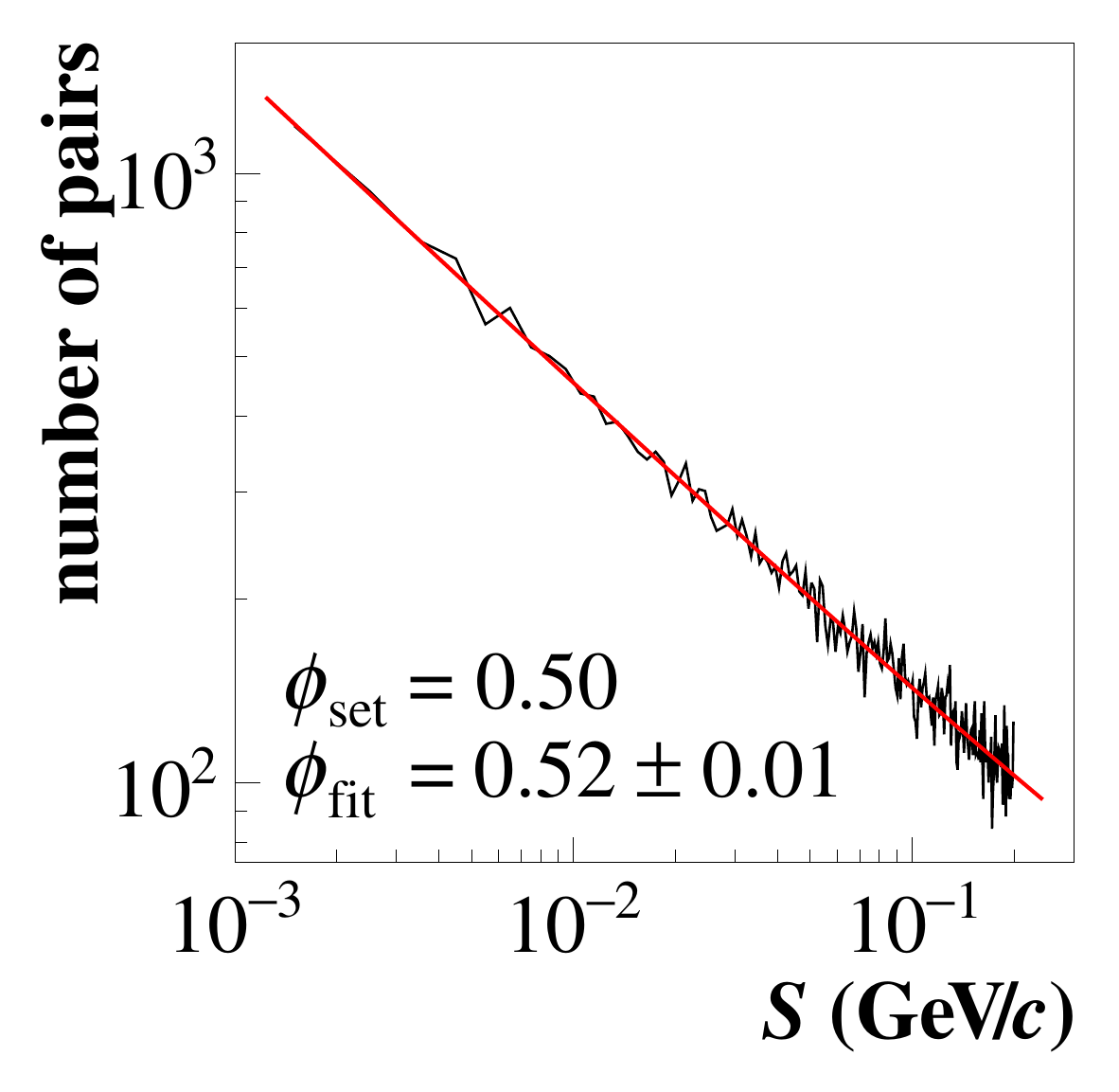}\hfill
  \includegraphics[width=.33\textwidth]{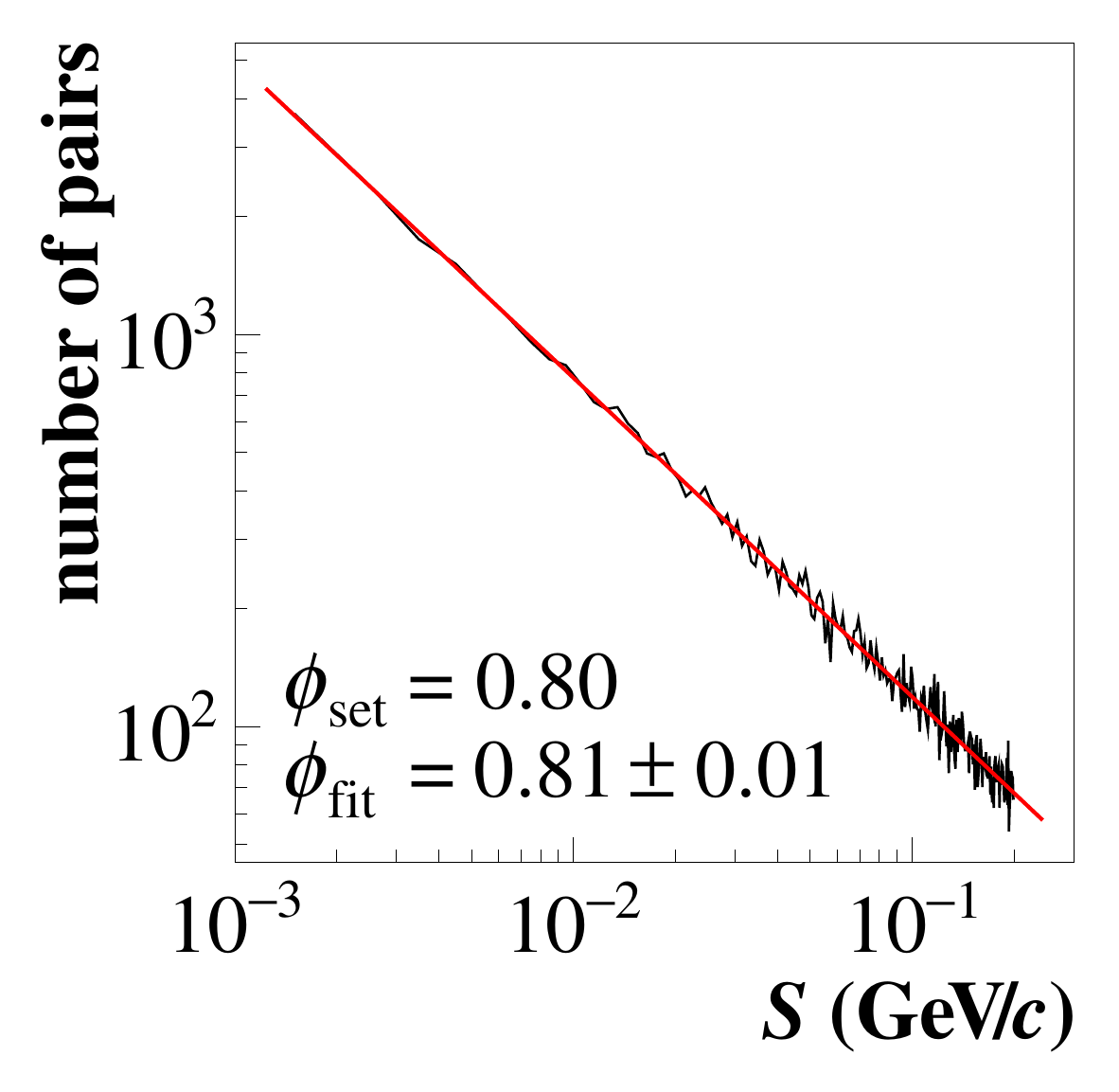}
  \caption{
    Illustration of the model performance for three sample settings (details in text).
    \emph{Top}: distributions of single-particle transverse momenta of correlated (black line) and uncorrelated (blue line) particles. \emph{Bottom}: distributions of average transverse momentum differences $S$ of correlated particle pairs generated with $\phi_\text{set}$ (black line) fitted with a power-law with $\phi_\text{fit}$ exponent (red line).
  }
  \label{fig:dist}
\end{figure}

The model's key feature is introducing a given power-law correlation of particles while preserving
a given single-particle transverse momentum distribution. To test it, 10\textsuperscript{5} events with $N=4$ particles (two correlated and two uncorrelated) with different model's settings have been generated:
\begin{enumerate}
    \item $\rho_{\pt}(\pt) = \pt e^{-6\pt}$ and $\phi_\text{set}=0.65$ (default model's values),
    \item $\rho_{\pt}(\pt) = e^{-3\pt}$ and $\phi_\text{set}=0.50$,
    \item $\rho_{\pt}(\pt) = 1-\pt/1.5$ and $\phi_\text{set}=0.80$.
\end{enumerate}
The relevant distributions for produced particles are shown in Fig.~\ref{fig:dist}. Histograms of single-particle transverse momenta (\emph{top}) for correlated (black line) and uncorrelated (blue line) particles are almost identical and both follow requested $\rho_{\pt}(\pt)$ distributions. Histograms of average transverse momentum differences $S$ (\emph{bottom}) of correlated particle pairs (black line) fitted with a power-law (red line) shows that the obtained value of $\phi_\text{fit}$ exponent agrees with the input value $\phi_\text{set}$ for all three cases.

Generating these data sets took approximately 200~ms each.

\subsection{Scaled Factorial Moments for the model data}
\label{sec:sfm}

The model's main purpose is to study SFM, which is used within intermittency analysis as a tool for locating the CP. Therefore, it must generate particles with properties expected for the CP.
One of the properties, power-law correlation, is explicitly built into the model. Within the framework of intermittency analysis, this correlation results in a power-law dependence of SFM of the order $q$ with respect to the number of (2D) $p_\mathrm{x}$--$p_\mathrm{y}$ cells $M^{2}$:
\begin{equation}
  \Delta F_{q}(M) \propto (M^{2})^{\varphi_{q}}\,,
\end{equation}
where the model power-law exponent $\phi$ is related to the second intermittency index $\varphi_{2}$ as
\begin{equation}\label{eq:varphi2}
  \varphi_{2} = \frac{\phi + 1}{2}\,.
\end{equation}

Additionally, the exponents (intermittency indices) should obey a linear relation:
\begin{equation}\label{eq:varphiq}
  \varphi_{q} = (q - 1)\cdot \varphi_{2}\,.
\end{equation}
This relation follows from the definition of $S$ and the dimensionality of the transverse momentum space used in the intermittency analysis.

\begin{figure}[!htb]
  \centering
  \includegraphics[width=0.5\textwidth]{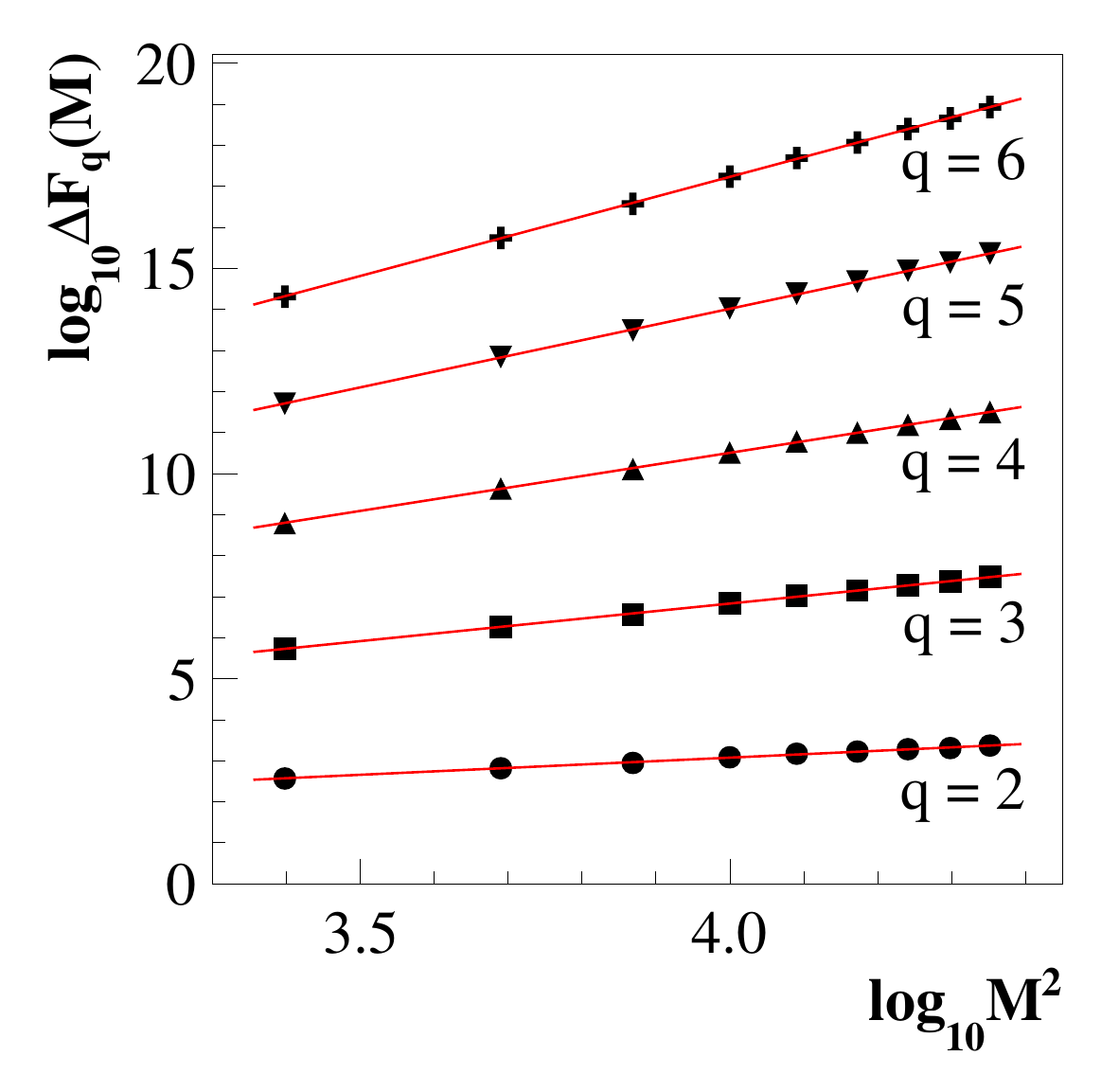}\hfill
  \includegraphics[width=0.5\textwidth]{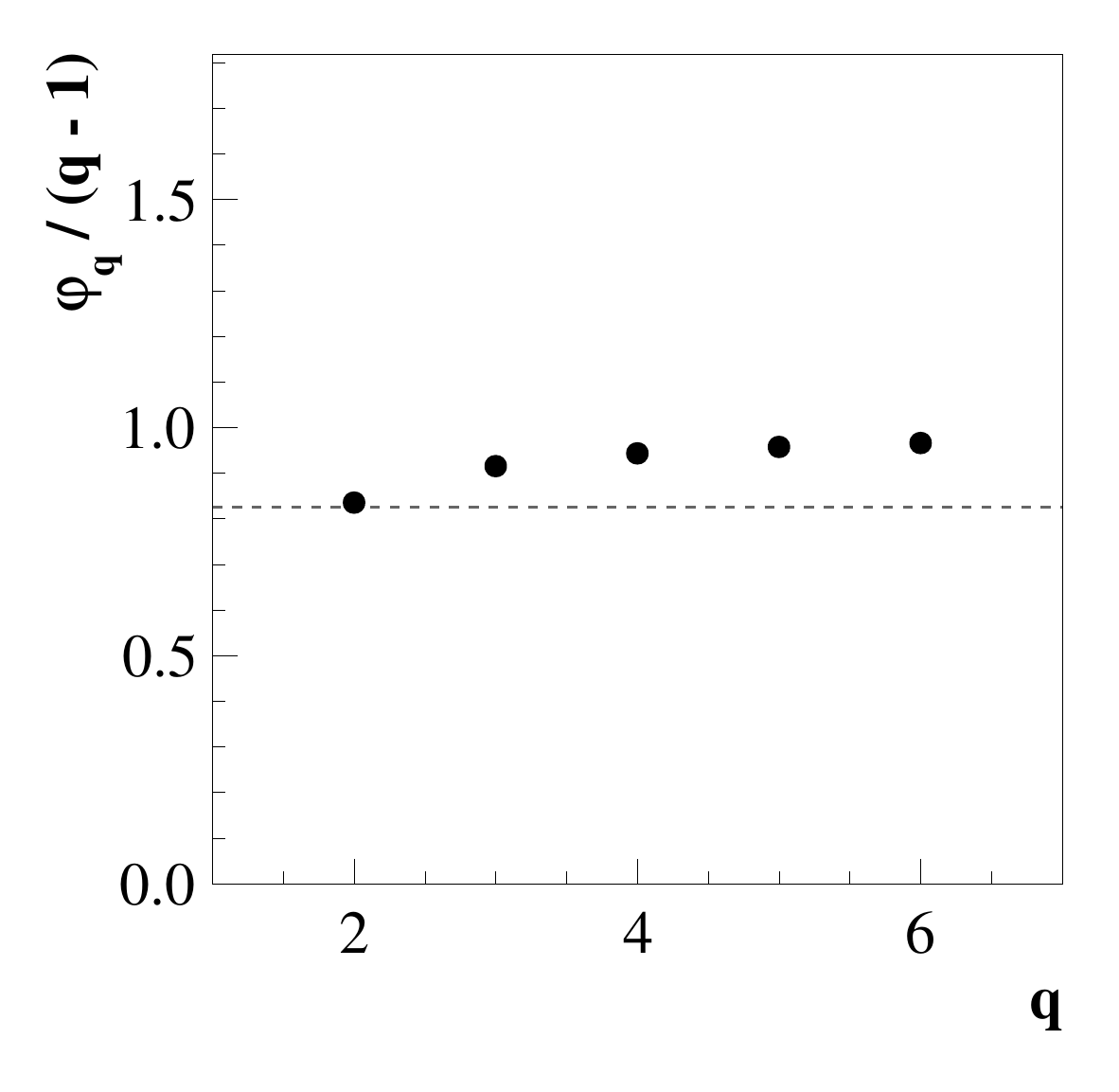}\\
  \includegraphics[width=0.5\textwidth]{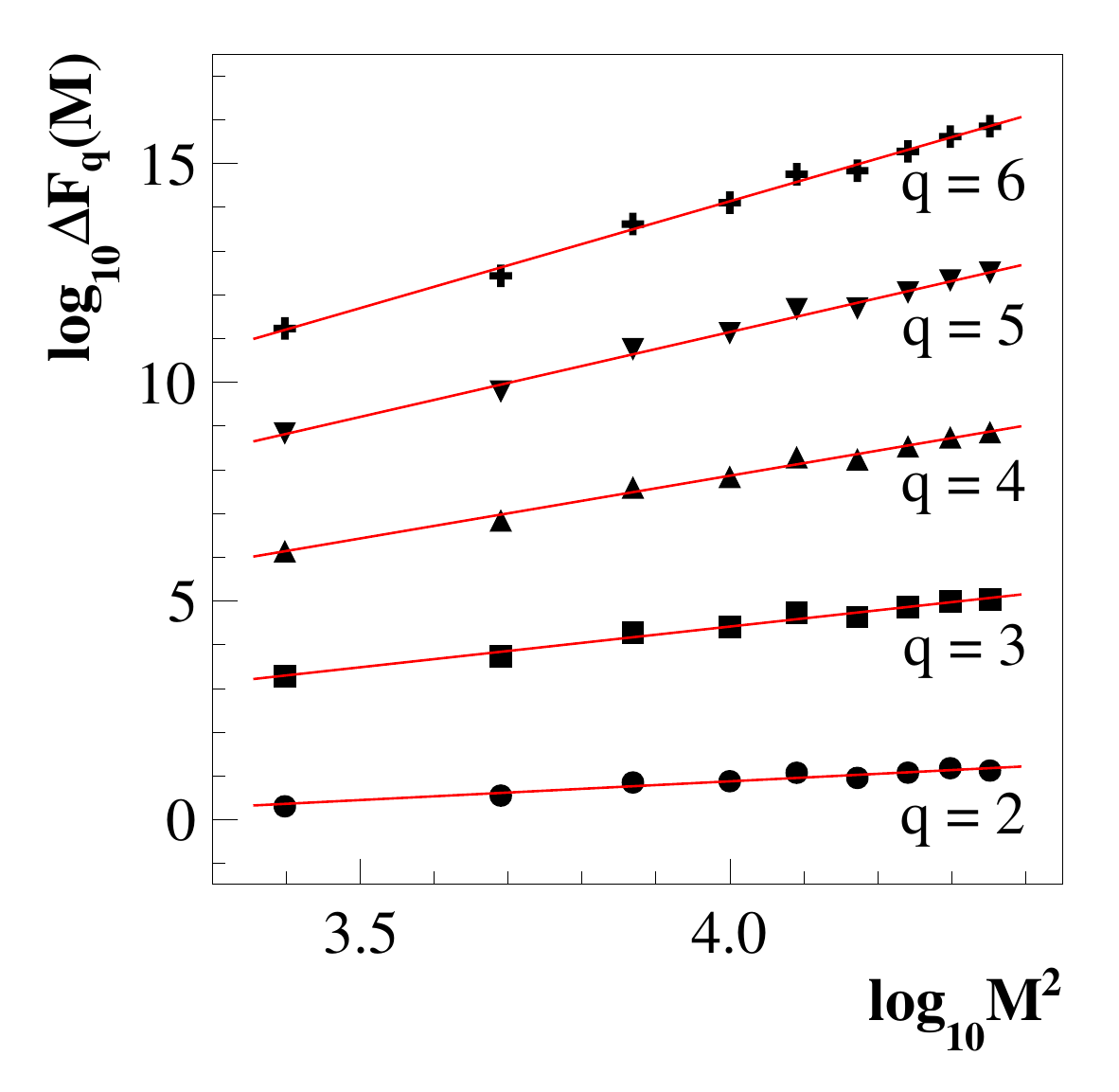}\hfill
  \includegraphics[width=0.5\textwidth]{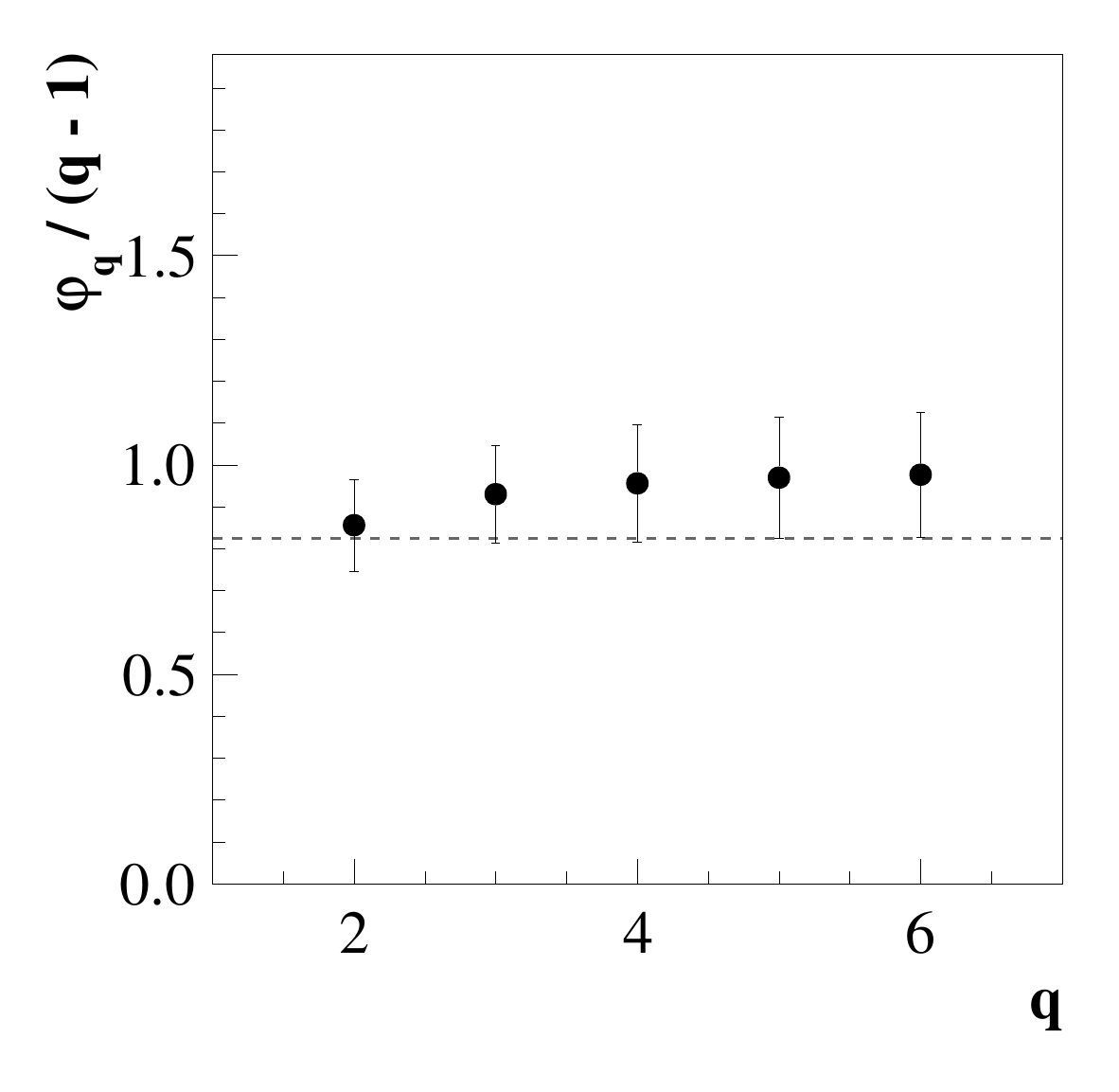}
  \caption{
    Illustration of intermittency results obtained from the model data. Two examples of 100k events generated using the default transverse momentum distribution and correlation exponent ($\phi=0.65$). \emph{Top}: each event contains only six correlated particles ($N=6$). \emph{Bottom}: Gaussian ($\langle N \rangle = 10$, $\sigma_{N}=2$) multiplicity with only 1\% of particles correlated.
    \emph{Left}: Scaled Factorial Moments of the orders from $q=2$ to $q=6$ (black markers) and their power-law fits shown as straight red lines in log-log scale.
    \emph{Right}: corresponding intermittency indices $\varphi_{2},\dots,\varphi_{6}$ (power-law exponents obtained from SFM fits) scaled with SFM order with expected value ($\varphi_{q}/(q-1) = 0.825$, see Eqs.~\ref{eq:varphi2},~\ref{eq:varphiq}) marked as dashed line.  
  }
  \label{fig:sfm}
\end{figure}

To test these features, two sets of 10\textsuperscript{5} events, have been generated using groups of six particles ($g=6$) correlated with $\phi=0.65$:
\begin{enumerate}
    \item each event containing exactly six correlated particles ($N=6$) and no uncorrelated particles ($r_{1}=1$), and
    \item using Gaussian multiplicity ($\langle N \rangle = 10$, $\sigma_{N}=2$) with only 1\% of particles being correlated ($r_{1}=0.01$).
\end{enumerate}
Results are shown in Fig.~\ref{fig:sfm}. SFMs up to $q=6$ order (black markers) fitted with power-law (red lines) in log-log scale (\emph{left}) manifest good fit for all orders for both data sets. The obtained exponents $\varphi_{2},\dots,\varphi_{6}$ scaled by $(q-1)$ (\emph{right}) for better visualization of the expected linear behavior characteristic of self-similar fluctuations.

\section{Summary and outlook}
\label{sec:summary}

This work is motivated by experimental searches for the critical point of the strongly interacting matter in heavy-ion collisions. A model introducing a power-law correlation predicted near the CP was presented. It does not aim to describe the microscopic dynamics of particle production near the critical point, but rather provides a controlled environment for testing analysis techniques. The expected scaling behavior of $F_{q}(M)$ in $M^{2}$, as well as a linear relation of obtained intermittency indices, is observed. Introducing correlations between particles does not affect transverse momentum and multiplicity distributions.

The model can be used to study the impact of detector effects (e.g. acceptance, efficiency, resolution, etc.) on the behavior of the scaled factorial moments. The simplicity and computational speed of the model make it particularly suitable for large-scale detector-response and acceptance studies.

The source code can be made available upon request.

The author would like to thank Marek Ga\'zdzicki for the motivation, help and critical comments.

This work was supported by the Polish National Science Centre grant 2018\slash 30\slash A\slash ST2\slash 00226.

\bibliographystyle{elsarticle-num}
\bibliography{references}

\end{document}